\documentclass[a4paper,reqno]{amsart}
\usepackage{geometry}
\usepackage{fullpage}
\usepackage[latin1]{inputenc}
\usepackage[T1]{fontenc}
\usepackage{amsfonts}
\usepackage{amssymb}
\usepackage{amsmath}
\usepackage{amsthm}
\usepackage{graphicx}
\usepackage[dvipsnames]{xcolor}
\usepackage{afterpage}
\usepackage[colorlinks=true, linkcolor=magenta, citecolor=cyan, urlcolor=green]{hyperref} 
\usepackage{bbm}
\usepackage{latexsym}
\usepackage{amsaddr}

\numberwithin{equation}{section}

\definecolor{light}{gray}{.9}

\def\be{\begin{equation}}
\def\ee{\end{equation}}
\def\bea{\begin{eqnarray}}
\def\eea{\end{eqnarray}}

\def\d{\delta}
\def\g{\gamma}

\newcommand{\meanv}[1]{\left\langle#1\right\rangle}

\newcommand{\nocontentsline}[3]{}
\newcommand{\tocless}[2]{\bgroup\let\addcontentsline=\nocontentsline#1{#2}\egroup}

\DeclareMathSymbol{\leqslant}{\mathalpha}{AMSa}{"36} \DeclareMathSymbol{\geqslant}{\mathalpha}{AMSa}{"3E} 
\DeclareMathSymbol{\eset}{\mathalpha}{AMSb}{"3F}    
\renewcommand{\leq}{\;\leqslant\;}                   
\renewcommand{\geq}{\;\geqslant\;}

\title{Disentangling group and link persistence in Dynamic Stochastic block models}
\date{\today}

\author{P. Barucca}
\address{Department of Computer Science, University College London, UK}
\author{F. Lillo}
\address{Department of Mathematics, University of Bologna, Italy}
\author{P. Mazzarisi, D.Tantari}
\address{Scuola Normale Superiore, Pisa, Italy}

\begin{document}
\maketitle

\begin{abstract}
We study the inference of a model of dynamic networks in which both communities and links keep memory of previous network states. By considering maximum likelihood inference from single snapshot observations of the network, we show that link persistence makes the inference of communities harder, decreasing the detectability threshold, while community persistence tends to make it easier. We analytically show that communities inferred from single network snapshot can share a maximum overlap with the underlying communities of a specific previous instant in time. This leads to time-lagged inference: the identification of past communities rather than present ones. Finally we compute the time lag and propose a corrected algorithm, the Lagged Snapshot Dynamic (LSD) algorithm, for community detection in dynamic networks. We analytically and numerically characterize the detectability transitions of such algorithm as a function of the memory parameters of the model and we make a comparison with a full dynamic inference. 
\end{abstract}

\vspace{1cm}

Community detection in time-evolving interacting systems is an open problem in data mining. Temporal networks \cite{holme2012temporal} provide a framework to study the dynamic evolution of interacting systems, and can be analysed as a sequence of  network snapshots. In this paper we study the problem of learning the dynamic evolution of the community structure of a temporal network with link and community persistence. The problem of disentangling the two different  sources of persistence in a model of temporal networks has been recently addressed in \cite{emid} in a  slightly different context concerning  Dynamic Fitness Models.
On the other hand community detection is a long-standing problem that has been thoroughly studied in the static network case with various approaches: modularity maximization \cite{newman2016community}, spectral methods \cite{hendrickson1995improved,krzakala2013spectral}, belief-propagation \cite{decelle2011asymptotic}, and other heuristic algorithms \cite{blondel2008fast}.  

In this paper we focus on Stochastic Block models with dynamic community structure and link persistence, which introduce time correlations in the network structure. When time correlations are present, the information obtained from the inference on individual snapshots might be contaminated by the past history of the system. This is analogous to what happens in multilayer networks \cite{boccaletti}, for which the analysis cannot be decomposed into the separate analysis over each layer if they are correlated.

Static stochastic block models have been shown to display a detectability transition \cite{decelle2011asymptotic,mossel2013proof1,mossel2013proof2} when the ratio between the average degree within a block of nodes and the average degree towards different blocks, i.e. the assortativity parameter, becomes too low: below a critical value of assortativity, detection becomes computationally hard. 

Recently the problem was also investigated in temporal networks \cite{mucha2010community,yang2011detecting,bassett2013robust,bazzi2016community} and in a specific case of Markovian community structures \cite{ghasemian2016detectability}. In this latter dynamic network model, it was shown that persistence in communities can help detection, by decreasing the detectability threshold: a weaker assortativity is required to infer communities with respect to the static case. On the contrary, we show that persistence in relations can hinder detection, eventually causing the detection of old communities instead of the ones present at the time the detection is performed. We compute analytically the time lag in community detection and provide a dynamic community detection algorithms for the model under study. The method is built upon optimal static algorithms on individual snapshots combined with our analytic result to correct for the time lag.

The paper is organized as follows: in section \ref{sec:1} we define the dynamic stochastic block model where both communities and links are persistent in time. In section \ref{sec:2} we study the single snapshot inference and we show how link persistence leads to time lagged inference, that is the detection of past communities rather than present ones. In Section \ref{sec:3} we introduce the lagged snapshot dynamic (LSD) algorithm, that corrects static detection algorithm for the time lag. In Section \ref{sec:4} we show that the  LSD algorithm can be considered an interesting tradeoff between the accuracy of a full dynamic inference and the simplicity of a naive single snapshot inference.   We conclude suggesting new possible  directions of research in Bayesian inference for temporal networks.

\section{Definition of the model}\label{sec:1}

We consider a Dynamic Stochastic Block Model (DSBM) with link persistence, i.e. at each time step the presence of a link between two nodes is copied from the previous time with probability $\xi$, while with probability $1-\xi$ the link is generated according to a SBM where the community structure changes over time.  Several models of DSBM were previously introduced for community detection in dynamic networks \cite{zhang2016random,xu2014dynamic,xu2015stochastic,ghasemian2016detectability}. Our variant includes both link and community persistence. 
The SBM is a classical generative model for static networks with community structure, where a network $(V,\boldsymbol{A})$ with $|V|=N$ nodes and adjacency matrix $\boldsymbol{A}$ is generated as follows. According to a prior $\{q_r\}_{r=1}^k$ over $k$ possible choices, each node $i\in V$ is assigned to a community $g_i$ with probability $q_{g_i}$. Edges are then generated according to a  $k\times k$ affinity matrix $\boldsymbol{p}$ and the community structure $\boldsymbol{g}$: each couple of nodes $i,j \in V$ are linked independently with probability $p_{g_ig_j}$.

\begin{figure}
\includegraphics[scale=0.5]{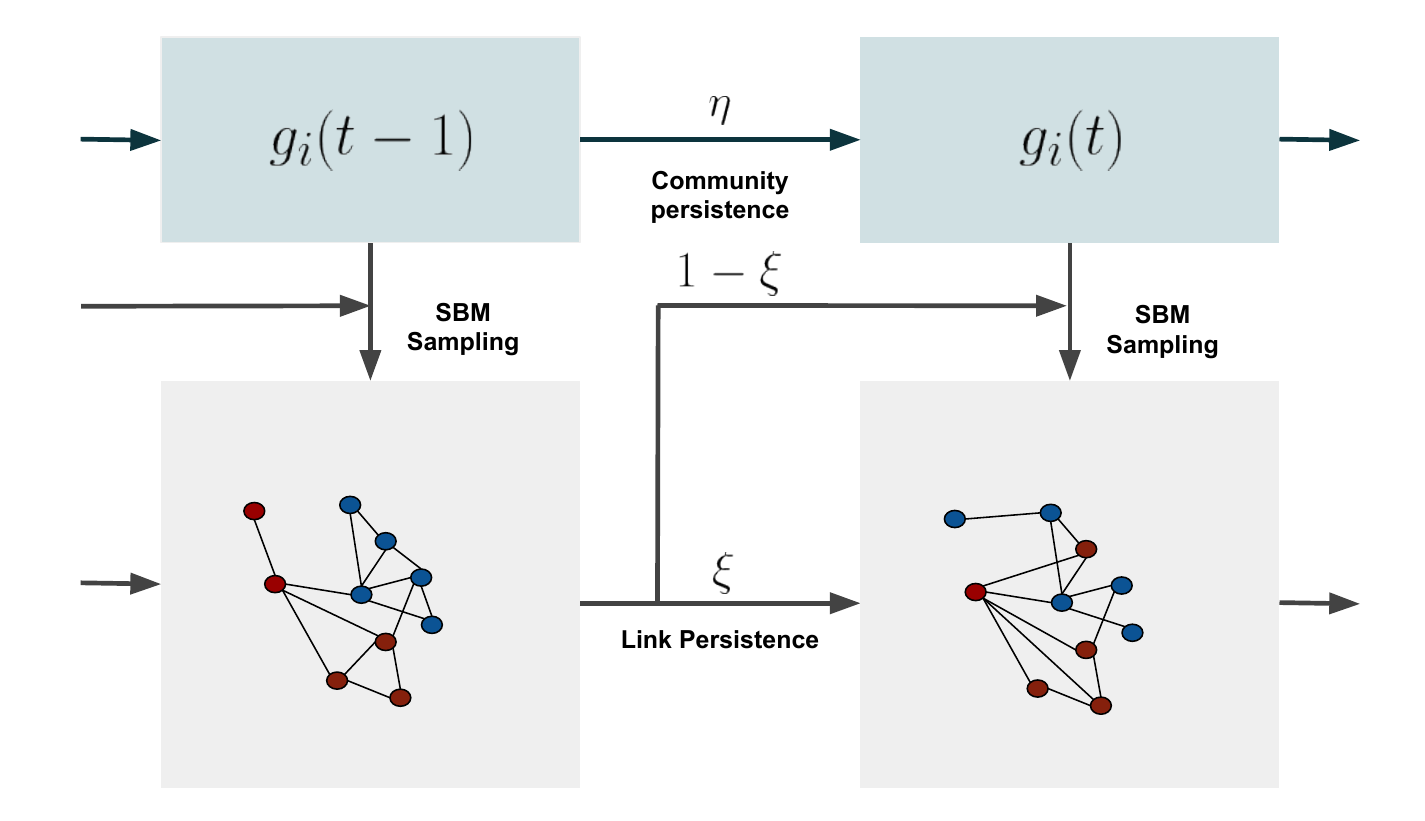}
\caption{Schematic representation of the persistent dynamic blockmodel. Label is kept with probability $\eta$ and randomly changed with probability $1-\eta$, while the link relation, i.e. the presence or absence of a link, is conserved with probability $\xi$ or resampled with the new block structure with probability $1-\xi$.}\label{fig:flow}
\end{figure}

In the DSBM the community structure changes over time. It consists of a sequence of networks $(V,\boldsymbol{A}^t)_{t=1}^T$, each with its own  community structure $\boldsymbol{g}^t$.  We will indicate with $\boldsymbol{A}=\{\boldsymbol{A}^0,...,\boldsymbol{A}^T\}$ the sequence of observed adjacency matrices and with $\boldsymbol{g}=\{\boldsymbol{g}^0,...,\boldsymbol{g}^T\}$ the sequence of community structures. As in \cite{ghasemian2016detectability}, the dynamic of each node's assignment $g_i^t$ is an independent Markov process with transition probability  
$
P(g_i^t |g_i^{t-1})=\eta\  \d_{g_i^t,g_i^{(t-1)}}+(1-\eta) q_{g_i^t},
$
meaning that  with probability $\eta$ a node remains in the same community, otherwise it changes randomly to a group $r$ (including $g_i^{t-1}$) with probability $q_r$. Since at $t=0$  labels are assigned according to the prior, it is
\be
P(\boldsymbol{g})= \prod_{i=1}^N \left [ \prod_{t=1}^T \eta\  \d_{g_i^t,g_i^{(t-1)}}+(1-\eta) q_{g_i^t} \right] q_{g_i^0}
\ee   
Adding link persistence to the DSBM we obtain the persistent dynamic model, see the flow in Fig. \ref{fig:flow} 
\begin{eqnarray}
P(\boldsymbol{A}|\boldsymbol{g})&=&\prod_{(ij)}^N p_{g^0_ig^0_j}^{A^0_{ij}}(1-p_{g^0_ig^0_j})^{1-A^0_{ij}}  \times \\
&\times& \prod_{t=1}^T \left(\xi\  \d_{A_{ij}^t,A_{ij}^{(t-1)}} + (1-\xi) p_{g^t_ig^t_j}^{A^t_{ij}}(1-p_{g^t_ig^t_j})^{1-A^t_{ij}}\right), \nonumber
\end{eqnarray}
where the network at $t=0$ is generated according to a static SBM from $\boldsymbol{g}^0$. 
Thus the two parameters $\eta$ and $\xi$ can be interpreted as, respectively, the persistence of communities and the persistence of links. Community persistence models the tendency of nodes to remain in the same group over time. Link persistence models the preference of nodes in keeping pre-existent relations over time, for example because of the cost of adding or removing links in socio-economic networks \cite{amaral2000classes}.

Here we focus on the common choices of a uniform prior, i.e. $q_r=1/k$ $\forall r=1,\ldots , k$, and affinity matrix with a constant $p_{in}$ on the diagonal and another constant $p_{out}\le p_{in} $ off diagonal, the so called assortative \textit{planted partition model}  that is widely used as benchmark in the mathematics and computer science community detection literature \cite{decelle2011asymptotic,krzakala2013spectral,pp1,pp2}. Moreover we measure the level of assortativity  with a parameter $a\in[0,1]$ such that
\be\label{eq:pab}
\boldsymbol{p}= a\ k\bar{p} \mathbb{I} + (1-a) \bar{p} \boldsymbol{1}
\ee
interpolating between a fully assortative $k\bar{p} \mathbb{I}$ (proportional to the identity matrix) and a fully random $\bar{p} \boldsymbol{1}$ (proportional to a matrix of ones) affinity matrix, with fixed mean degree $N/k^2 \sum_{ab} p_{ab}=N\bar{p}$. We are interested in the sparse regime $\bar{p}=\bar{c}/N$, that is the most challenging from the inference perspective, since most of real networks of interest are sparse and because sparsity allows to carry out asymptotically optimal analysis.

The central problem is to study under which conditions we can detect, better than chance, the correct  labeling of the latent communities $\boldsymbol{g}$ from the observation of $\boldsymbol{A}$, together with the most probable model's parameters $\phi=(a,\xi,\eta)$. For the static SBM, it was shown (and proved at least for $k=2$  \cite{mossel}) that there exists a sharp threshold below which no algorithm can perform better than chance in recovering the planted community structure. This threshold occurs, in terms of the parametrization $(\ref{eq:pab})$ at $a=a^c:=\bar{c}^{-1/2}$
meaning that there is a necessary minimum signal to noise ratio, in terms of assortativity, under which a community structure may still exist but is undetectable.
The Bayesian inference approach considers the posterior distribution of the latent assignments
\be\label{eq:l2}
P(\boldsymbol{g}|\boldsymbol{A},\phi)=\frac{P(\boldsymbol{A},\boldsymbol{g}|\phi)}{\sum_{\boldsymbol{\g}}P(\boldsymbol{A},\boldsymbol{\g}|\phi)}=Z^{-1}e^{-\mathcal{H}(\boldsymbol{g}; \boldsymbol{A},\phi)},
\ee 
where we have defined $\mathcal{H}(\boldsymbol{g}; \boldsymbol{A},\phi)=-\log P(\boldsymbol{A},\boldsymbol{g}|\phi)$, for inferring a set of statistically significant communities $\hat{\boldsymbol{g}}$ and  the posterior distribution over the model parameters
\be\label{eq:l1}
P(\phi|\boldsymbol{A})=\frac{P(\phi)}{P(\boldsymbol{A})} \sum_{\boldsymbol{\g}}P(\boldsymbol{A},\boldsymbol{\g}|\phi)\propto P(\phi) Z 
\ee
to learn the most likely set of parameters $\hat{\phi}$ given the data. Using smooth priors $P(\phi)$, $\hat{\phi}$ is obtained by maximizing  the likelihood $(\ref{eq:l1})$ with respect to $\phi$, i.e. by solving the equations
\be\label{eq:derphi}
\sum_{\boldsymbol{g}} Z^{-1}e^{-\mathcal{H}(\boldsymbol{g}; \boldsymbol{A},\phi)} \partial_\phi \mathcal{H} (\boldsymbol{g}; \boldsymbol{A},\phi) =\meanv{\partial_\phi \mathcal{H} (\boldsymbol{g}; \boldsymbol{A},\phi)}  =0.
\ee
Since the maximization of the likelihood  ($\ref{eq:l1}$) requires computing expectations w.r.t the posterior ($\ref{eq:l2}$), this is called Expectation-Maximization (EM) procedure \cite{fridman2001elements}. The criticality of this approach is in the summation over all possible assignments whose number grows exponentially with $N$. Overtaking this problem is usually done by Monte Carlo (MC) sampling \cite{peixoto2013parsimonious} or  by using belief propagation (BP) algorithms \cite{decelle2011asymptotic,decelle2011inference}. Both provide  an estimate of the posterior $(\ref{eq:l2})$ in terms of their marginals. From them, a partition is obtained by assigning each node to its most likely group 
$$\hat{g}_i= \operatorname{argmax}_r P(g_i=r|\boldsymbol{A},\hat{\phi}).$$
This is known \cite{iba} to be an optimal estimator, maximising the overlap with the planted assignment
\be\label{eq:overlap}
q(\boldsymbol{g},\boldsymbol{\hat{g}})=\frac{N^{-1}\sum_{i=1}^N \d_{\hat{g}_i g_i} - \max_r q_r}{1-\max_r q_r},
\ee
where the normalization is chosen to ensure $q=0$ if labels are assigned randomly.

In a static network, generated by a static SBM,  the EM procedure described before provides a set of inferred assignments $\hat{\boldsymbol{g}}$ together with an estimate of the affinity matrix $\hat{p}_{ab}$, obtained using the static posterior corresponding to $\mathcal{H}(\boldsymbol{g}; \boldsymbol{A},\boldsymbol{p})=-\log p_{g_ig_j}^{A_{ij}}(1-p_{g_ig_j})^{1-A_{ij}} $ and solving iteratively the equation
\be
p_{ab}=\meanv{ \frac{\sum_{(i,j)} A_{ij}\d_{g_i,a}\d_{g_j,b} }{\sum_{(i,j)} \d_{g_i,a}\d_{g_j,b} }}.
\ee
This is the equivalent of Eq. $(\ref{eq:derphi})$, by deriving w.r.t. $p_{ab}$. The value of $\hat{a}$ is obtained by fitting it on the inferred affinity matrix as in Eq. (\ref{eq:pab}). Finally, throughout the paper,  we will use the static EM procedure introduced in \cite{decelle2011asymptotic} where a BP algorithm is used for the expectation step, i.e. the estimate of the  posterior marginals.

\section{Single snapshot inference}\label{sec:2}

\begin{figure}
\includegraphics[scale=0.6]{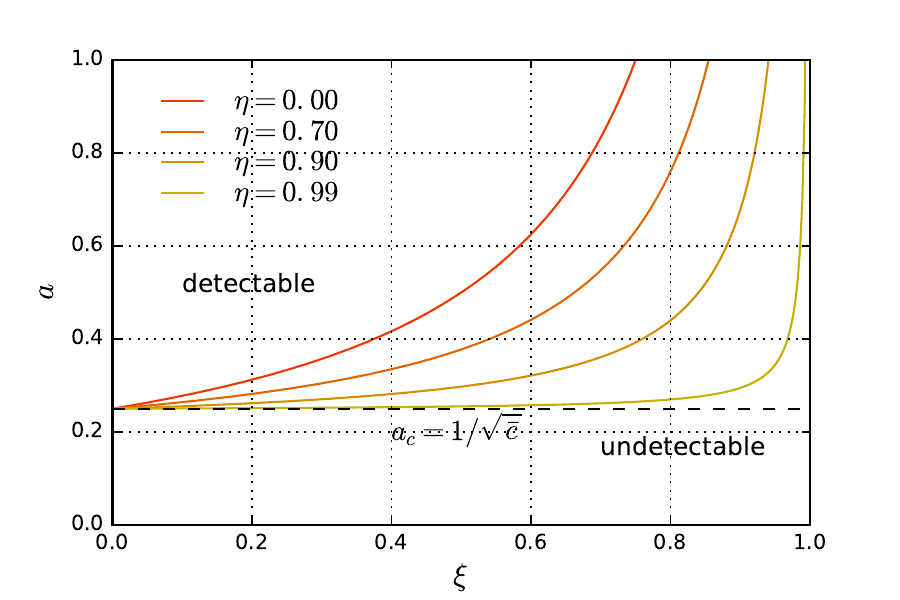}
\caption{Asymptotic phase space for single snapshot detectability as function of assortativity $a$ and community/ link persistences, $\eta$ and $\xi$, compared to the static threshold $a_c$, in the case of two equally sized planted groups $k=2$. }\label{fig:phdiag}
\end{figure}

The inference for the full dynamical model is complicated by the presence of both the link and community persistence. Here we ask first which community structure is inferred from a single snapshot of the dynamic network at a time $t$. This might occur, for example, if one is unaware that $\boldsymbol{A}^t$ is one observation of a dynamic process.
Thus we need to compute the posterior $P(\boldsymbol{g}^t|\boldsymbol{A}^t)$ giving the probability of community structure when only the information on the network at time $t$ is used. It holds that the posterior $P(\boldsymbol{g}^t|\boldsymbol{A}^t)$ is that of a static SBM with an effective assortativity 
\be\label{eq:at}
a^{t}_{\xi,\eta}=a \  \epsilon^t_{\xi,\eta}= a\left(\frac{1-\xi}{1-\xi\eta^2} + (\xi\eta^2)^t\frac{\xi(1-\eta^2)}{1-\xi\eta^2} \right).
\ee

In fact it is sufficient to note that, from Bayes' rule, $P(\boldsymbol{g}^t|\boldsymbol{A}^t)\propto P(\boldsymbol{A}^t|\boldsymbol{g}^t)$, that can be always be written as
\be
P(\boldsymbol{A}^t|\boldsymbol{g}^t)=\prod_{(i,j)} (p^{t}_{g^t_ig^t_j})^{A^t_{ij}}{(1-p^{t}_{g^t_ig^t_j})}^{1-A^t_{ij}},
\ee
with $p^{t}_{ab}:=P(A^t_{ij}=1|g^t_i=a,g^t_j=b)$. Marginalising over previous network instances we get the recursive equation
\begin{eqnarray}
p^{t}_{ab}&=& \xi\  P(A^{t-1}_{ij}=1|g^t_i=a,g^t_j=b) + (1-\xi) p_{ab}.\nonumber\\
&=& \xi\ \left(\eta^2 p^{t-1}_{ab}+(1-\eta^2)\bar{p} \right)+ (1-\xi) p_{ab},
\end{eqnarray}
where in the first equality we have conditioned and summed over $A^{t-1}_{ij}$, while in the second over $g^{t-1}_i,g^{t-1}_j$ and where we used that $P(A^t_{ij}=1)=\bar{p}$ and $P(g^t_i=a)=1/k$ for every $i,j,t,a$,  that can be proved recursively.
Since $p^0_{ab}$ is simply $p_{ab}$ we get 
\be
p^{t}_{ab}=\left( \xi(1-\eta^2)\bar{p}+(1-\xi) p_{ab}\right)\sum_{\ell=0}^{t-1}\left(\xi\eta^2\right)^\ell +\left(\xi\eta^2\right)^tp_{ab},\nonumber
\ee
that gives $(\ref{eq:at})$ once used the representation $(\ref{eq:pab})$.

Eq. $(\ref{eq:at})$ states that the posterior of a single snapshot of a DSBM is equal to the posterior of a static SBM with modified assortativity parameter.  It is important to outline it does not imply that a single snapshot inference gives the planted assignments with modified assortativity parameter. Instead it states that, {\it if the inferred assignments are the planted ones}, then the estimated assortativity is the one of Eq. (\ref{eq:at}), i.e. $\hat a= a^{t}_{\xi,\eta}$, which is smaller than the value $a$ of the model. This happens because the link persistence $\xi$ decreases the effective assortative structure of the network, increasing the number of links assigned randomly with respect to those assigned on the base of their group labels. This effect is partially mitigated by the persistence of communities $\eta$ since it increases the probability that a link copied from a previous time is not actually random but was in turn assigned through the same community structure. 

One of the consequences of Eq. $(\ref{eq:at})$ is that the signal  provided by the observation of $\boldsymbol{A}^t$ to the community structure at the same time decreases  by the effect of the dynamics as $a^c \to a^c/\epsilon^t_{\xi,\eta}$,  reducing to the static one in absence of link persistence ($\xi=0$)\footnote{ Note that the detectability threshold from single snapshot is however higher than the threshold of the dynamic problem, i.e. the inference of all the assignments given the observation of the entire network series. For example  \cite{ghasemian2016detectability} considers a DSBM without link persistence and shows that the detectability threshold $a^c$ is in general lowered by the group persistence.}. For  $t\to \infty$, it is $a^\infty_{\xi,\eta}= a(1-\xi)/(1-\xi\eta^2)$.  Figure \ref{fig:phdiag} shows the asymptotic phase space as a function of $\xi$, where we have defined, in the same spirit of the static case, a  detectability line as $a(1-\xi)/(1-\xi\eta^2)=\bar c ^{-1/2}$.

\begin{figure}
\includegraphics[scale=0.5]{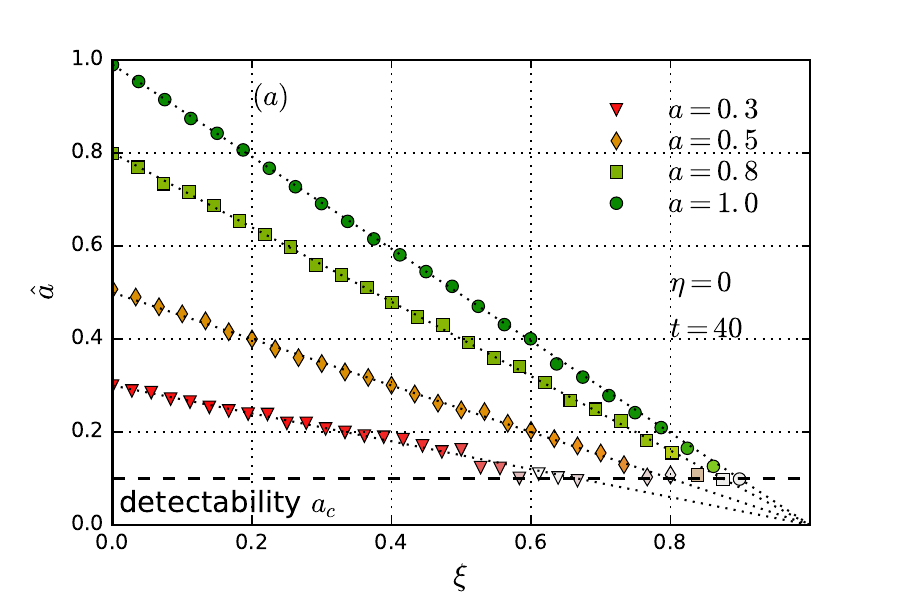}
\includegraphics[scale=0.5]{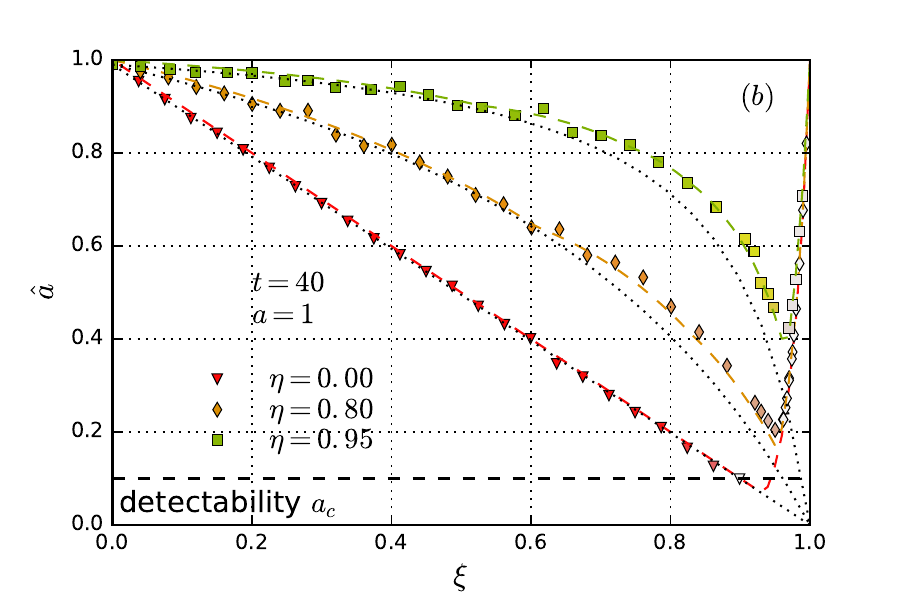}
\includegraphics[scale=0.5]{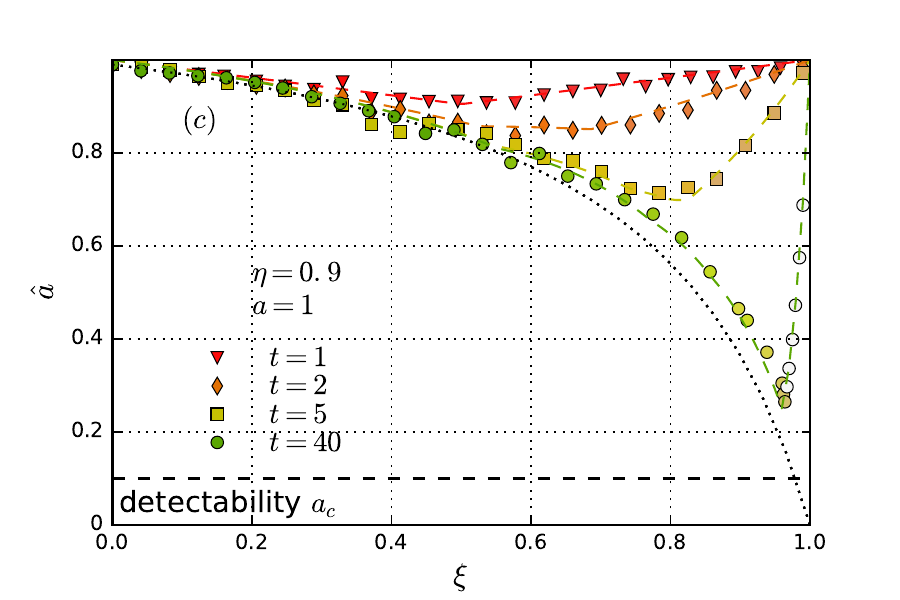}
\caption{Effective assortativity $\hat{a}$ inferred using static BP from single snapshot observations of a DSBM with link persistence, for different values of $\xi$, $\eta$ and $t$. Black dotted lines represent $a^t_{\xi,\eta}$ while colored dashed lines are the theoretical optimum $a^\star_t(\xi,\eta)$ of Eq. (\ref{eq:max}) . Each point is the result of the inference on a dynamic network with $N=300$, $T=40$, $\bar{c}=10$  and $k=2$ evolving communities. Vividness of colors is proportional to the overlap $q(\boldsymbol{g}^t,\hat{\boldsymbol{g}}^t)$ between the planted and the inferred communities. }\label{fig:astar}
\end{figure}

Figure \ref{fig:astar} compares the theoretical predictions of $\hat a$ with numerical simulations and BP inference of a DSBM, by varying planted parameters $(a,\xi,\eta,t)$ . In  panel (a) the agreement is very good and this holds also in the other panels in the regions when $\xi$ is small. However the panels (b) and (c) show that when both $\xi$ and $\eta$ are large, some discrepancies between the theoretical curve and the simulations appear.  This does not contradict necessarily Eq. $(\ref{eq:at})$, which gives the assortativity parameter if the inferred assignments are the planted ones (or at least close to them). We now show that indeed the observed discrepancies can be explained by the fact that, for large community and link persistence, the inferred assignments are closer to a {\it past} planted assignment than to those at the time when the single snapshot inference is performed.

\begin{figure}
\includegraphics[scale=0.6]{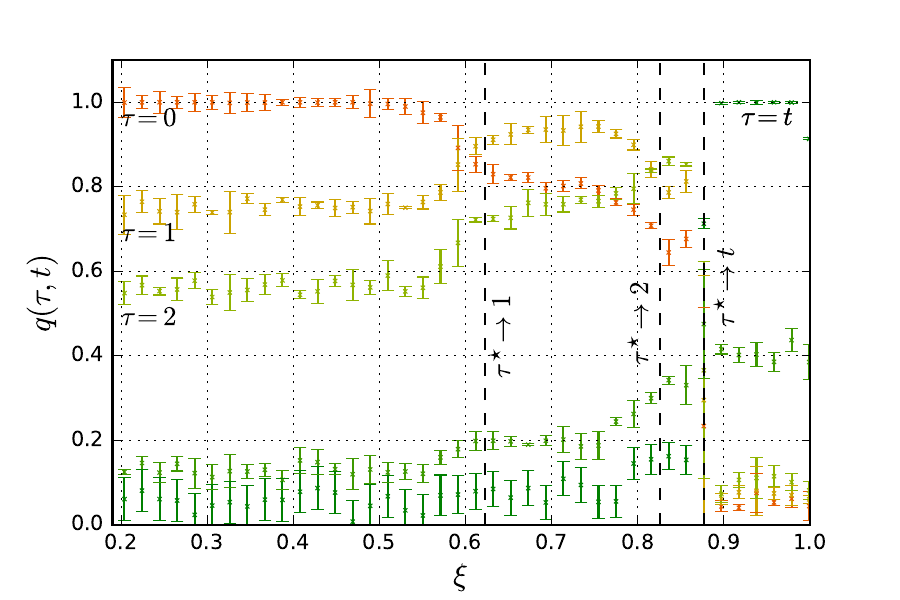}
\caption{Overlap $q(\hat{\boldsymbol{g}}^t,\boldsymbol{g}^{t-\tau})$ between the communities inferred at time $t$ $(t=10)$ and the planted assignment at time $t-\tau$ of a  dynamic network with $N=300$, $T=40$, $\bar{c}=10$  and $k=2$ evolving communities according to $\eta=0.75$.  Dashed lines are computed by solving the problem in Eq. (\ref{eq:max}). }\label{fig:overlap}
\end{figure}

Given a network sequence of length $T$ generated with parameters $(\xi,\eta,a)$, we call \textit{time lagged inference} the problem of inferring communities at time $t-\tau$ given the observation of the network at time $t$.  It holds that the posterior $P(\boldsymbol{g}^{t-\tau}|\boldsymbol{A}^t)\propto \prod_{ij}P(A^t_{ij}|\boldsymbol{g}^{t-\tau})$  is that of a static SBM with an effective assortativity
\be \label{th:lag}
a_{\xi,\eta}^{(t,\tau)}= \xi^\tau a_{\xi,\eta}^{t-\tau}+(1-\xi)\eta^2\frac{\eta^{2\tau}-\xi^\tau}{\eta^2-\xi} a,
\ee
where $a^{t}_{\xi,\eta}$ is given by $(\ref{eq:at})$.
In fact, as for Eq. $(\ref{eq:at})$ it sufficient to compute the quantity $\mathcal{L}^n=P(A^n_{ij}=1|\boldsymbol{g}^{t-\tau}=\boldsymbol{g})$, evaluated at $n=t$. For $n\geq t-\tau$, keepinig fixed $i,j$ and $t$ , it is
\begin{eqnarray}\label{eql}
\mathcal{L}^n&=&\sum_{A^{n-1}_{ij},\boldsymbol{g}^n} P(A^n_{ij}=1|\boldsymbol{g}^n,A_{ij}^{n-1}) P(\boldsymbol{g}^n,A_{ij}^{n-1}|\boldsymbol{g}^{t-\tau}=\boldsymbol{g})\nonumber\\
&=& \xi \mathcal{L}^{n-1}+(1-\xi) \sum_{\boldsymbol{g}^n} p_{g^n_ig^n_j} P(\boldsymbol{g}^n|\boldsymbol{g}^{t-\tau}=\boldsymbol{g}).
\end{eqnarray}
Moreover, defining $\mathcal{T}^n= \sum_{\boldsymbol{g}^n} p_{g^n_ig^n_j} P(\boldsymbol{g}^n|\boldsymbol{g}^{t-\tau}=\boldsymbol{g})$, for $n\geq t-\tau$ it is
\begin{eqnarray}\label{eqt}
\mathcal{T}^n&=& \sum_{\boldsymbol{g}^n,\boldsymbol{g}^{n-1}} p_{g^n_ig^n_j}P(\boldsymbol{g}^n|\boldsymbol{g}^{n-1})P(\boldsymbol{g}^{n-1}|\boldsymbol{g}^{t-\tau}=\boldsymbol{g})\nonumber\\
&=& \eta^2 \mathcal{T}^{n-1} +(1-\eta^2) \bar{p}
\end{eqnarray}
Solving $(\ref{eqt})$ and then  $(\ref{eql})$, i.e. the recursive equation $\mathcal{L}^n=\xi\mathcal{L}^{n-1}+(1-\xi) \mathcal{T}^n$ we get
\begin{eqnarray}
&&\mathcal{L}^t=\xi^\tau \mathcal{L}^{t-\tau}+(1-\xi)\sum_{\ell=0}^{\tau-1} \xi^\ell \mathcal{T}^{t-l}\nonumber\\
&=&\xi^\tau \mathcal{L}^{t-\tau}+(1-\xi)\sum_{\ell=0}^{\tau-1} \xi^\ell \left( \eta^{2(\tau-\ell)} p_{g_ig_j} +(1- \eta^{2(\tau-\ell)})\bar{p}\right)\nonumber.
\end{eqnarray}
Since $ \mathcal{L}^{t-\tau}$ corresponds to the non lagged $p^{t-\tau}_{g_ig_j}$ in Eq. $(\ref{eq:at})$,  we get Eq. $(\ref{th:lag})$
simply using the representation ($\ref{eq:pab}$).

\begin{figure}
\includegraphics[scale=0.5]{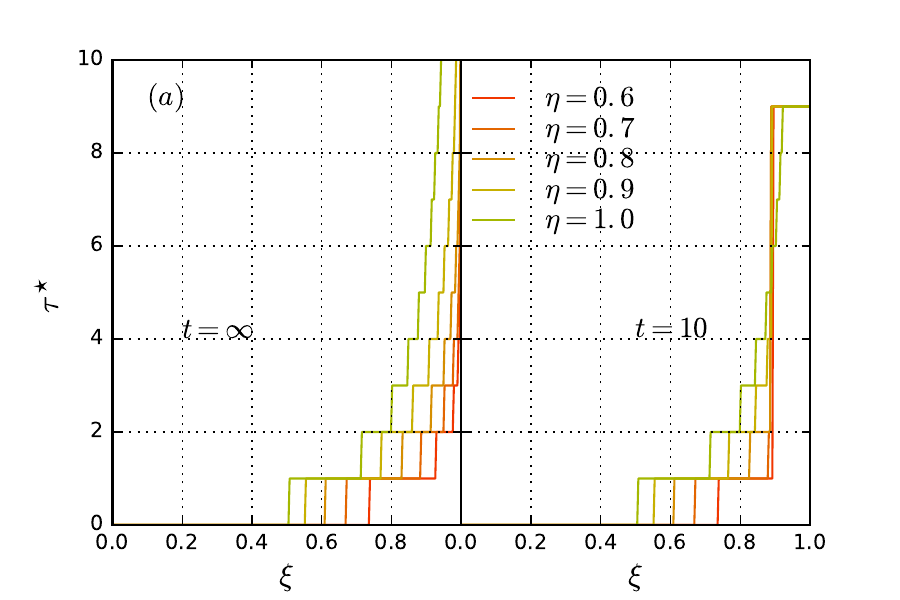}
\includegraphics[scale=0.5]{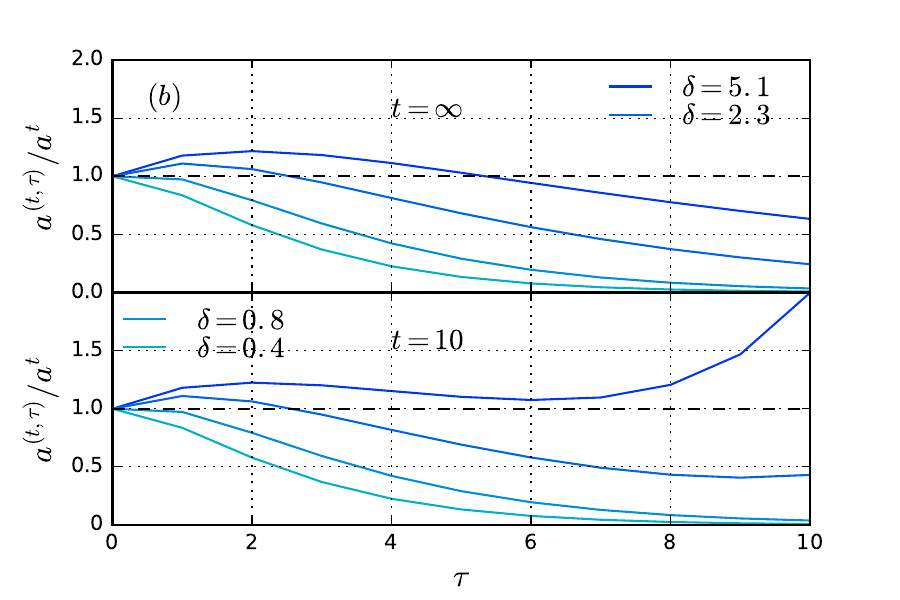}
\includegraphics[scale=0.5]{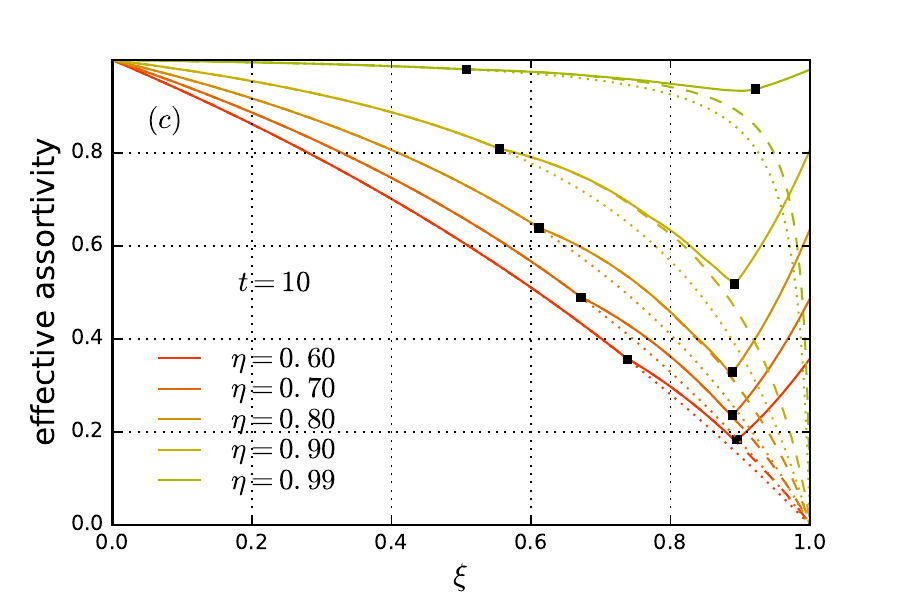}
\caption{ (a) Optimal lag $\tau^*_t(\xi,\eta)$ as function of $\xi$ and $\eta$ and  (b) $a^{(t,\tau)}_{\xi,\eta}/a^{t}_{\xi,\eta}$  as function of the lag $\tau$ for different  $\d=\eta^2 \xi /(1-\xi)$, both in the asymptotic limit $t\to\infty$  and for finite $t$. (c) Optimal effective assortativity $a^\star_t(\xi,\eta)$ (full lines) compared with the non lagged $a^t_{\xi,\eta}$ (dotted line) and the asymptotic $a^\star_\infty(\xi,\eta)$ (dashed lines). Black squares indicate the two transitions (see text). }\label{fig:Tstar}
\end{figure}

The meaning of Eq. $(\ref{th:lag})$ is that every lagged inference problem has the posterior of a static SBM with effective assortativity $a_{\xi,\eta}^{(t,\tau)}$. Thus fixing $t$ and varying $\tau$ we have a sequence of inference problems with the {\bf }same posterior, same input data $\boldsymbol{A}^t$, and only different effective assortativity, thus detectability threshold.  
Fig. \ref{fig:overlap} shows the overlap of Eq. (\ref{eq:overlap}) between the inferred communities $\hat{\boldsymbol{g}}^t$ and the planted ones at $t-\tau$ . For small $\xi$ the maximum overlap is with  ${\boldsymbol{g}}^{t}$, while for larger $\xi$ we observe a series of transitions where the largest overlap is with a ${\boldsymbol{g}}^{t-\tau}$ with $\tau>0$. We now show that the $\tau$ that maximizes the overlap $q(\hat{\boldsymbol{g}}^t,\boldsymbol{g}^{t-\tau})$ is the one for which the effective assortativity $a_{\xi,\eta}^{(t,\tau)}$ is maximal. To this end we define 
\be\label{eq:max}
a^{\star}_t (\xi,\eta)= \max_{\tau\leq t} a_{\xi,\eta}^{(t,\tau)};\ \ \ \ \ \ \ \ \ \tau^{\star}_t(\xi,\eta)=\underset{\tau\leq t}{\operatorname{argmax}} ~a_{\xi,\eta}^{(t,\tau)}
\ee 
Panel (a) of Fig. \ref{fig:Tstar} shows that for small link persistence $\xi$, $\tau^{\star}_t(\xi,\eta)=0$, i.e.  a single snapshot  inference solves the problem at the time of the observed snapshot $t$. At a critical $\xi$, depending on $\eta$ and $t$, it is $\tau^{\star}_t(\xi,\eta)>0$, suggesting that the inference procedure converges to the assignments at time $t-\tau^{\star}_t$. In fact the dashed lines in Fig. \ref{fig:overlap} are computed by solving the problem in Eq. (\ref{eq:max}) and it is clear that they correspond to the transitions in the overlap. Moreover the theoretical $a^\star_t(\xi,\eta)$ is shown in Fig. \ref{fig:astar} to be in perfect agreement with the inferred assortativity $\hat{a}$.

To get more intuition, we note that for large $t$ 
\be\label{eq:astar}
a_{\xi,\eta}^{(\tau)}:=\lim_{t\to\infty}a_{\xi,\eta}^{(t,\tau)}= a\left(\xi^\tau \frac{1-\xi}{1-\xi\eta^2}+(1-\xi)\eta^2\frac{\eta^{2\tau}-\xi^\tau}{\eta^2-\xi}\right).
\ee
Since $a_{\xi,\eta}^{(\tau)}\to 0$ as $\tau\to\infty$, when $a_{\xi,\eta}^{(1)}>a_{\xi,\eta}^{(0)}$, i.e. 
\be \label{eq:delta}
\delta:=\eta^2\frac{\xi}{1-\xi}>1
\ee
the maximum of $a_{\xi,\eta}^{(\tau)}$ is not anymore at $\tau=0$ (see the panel (b) of Fig. \ref{fig:Tstar}). 

For finite $t$, there is a finite size effect since the range of $\tau$ is bounded by $t$. In this situation for large $\xi$ and $\eta$ the maximum of $a^{(t,\tau)}_{\xi,\eta}$ is achieved at the extremum $\tau=t$ (panels (a) and (b) of Fig. \ref{fig:Tstar}). Finally, the  panel (c) of Fig. \ref{fig:Tstar} compares $a^\star_t(\xi,\eta)$, $a^t_{\xi,\eta}$, and $a^\star_\infty(\xi,\eta)$. The black squares indicate the two transitions, the first one from zero to positive $\tau^*$ (computed with Eq. (\ref{eq:delta}) )  and the second when $\tau^*=t$ due to the finite size effect. These correspond to the transitions observed in the empirical analysis of Fig. \ref{fig:astar}.

\section{Lagged snapshot dynamic (LSD) algorithm }\label{sec:3}

In this Section we propose a single snapshot  algorithm for the inference of the optimal assignments together with a set of learned model parameters  from the observation of a dynamic network. In Section \ref{sec:2} we showed how a naive single snapshot inference procedure, applied to a dynamic network with  group and link persistence,  introduces a systematic bias in the result. This bias takes the shape of a temporal lag: communities inferred at time $t$ share a maximum overlap with planted communities at time $t-\tau^\star$. This can affect also the goodness of the optimal parameters learned from data, for example the measured effective assortativity parameter is systematically overestimated at high link persistency.
For this reason we now propose a single snapshot algorithm able to detect and thus correct the possible presence of a temporal lag. Using only observations of the time series $\boldsymbol{A}^t$, we look for a set of inferred parameters $\hat{\eta}$, $\hat{\xi}$, $\hat{a}$ and group assignments $\hat{\boldsymbol{g}}^t$ using the following scheme, whose details are presented below:
\begin{enumerate}
\item  for each snapshot we estimate the assortativity and the assignments using a static method (e.g. BP on SBM);
\item we  estimate the link and group persistence $\hat{\xi}$ and $\hat{\eta}$ from the sequence of inferred assignments;
\item we compute the optimal lag $\hat \tau^*$ to get an unbiased estimation of the assortativity parameter and the correct assignments at time $t$ by considering the inferred assignments at time $t-\hat\tau^*$.
\end{enumerate}

We now detail the three phases of the LSD algorithm.

\subsection*{Single snapshot estimate.} For each snapshot observation $\boldsymbol{A}^t$ we perform the inference from a static SBM, as in \cite{decelle2011asymptotic}. The result is a set of assignment  $\boldsymbol{y}^t$ and an effective assortativity $\hat{a^\star}$. As shown in Section \ref{sec:2}, the use of a static procedure introduces a bias in the result: $\hat{a^\star}$  is a downward biased estimation of the assortativity parameter and $\boldsymbol{y}^t$ is an estimate of the planted assignment sequence but shifted by a lag $\tau^\star$, i.e. $\boldsymbol{y}^t=\hat{\boldsymbol{g}}^{t-\tau^\star}$. Clearly at this point $\tau^*$ is still unknown.

\subsection*{Estimate of the persistence parameters.} The inference of the persistence parameters $\hat{\xi}$ and $\hat{\eta}$ is performed by maximizing the likelihood ($\ref{eq:l1}$). Deriving the log-likelihood w.r.t. $\eta$ we get 
\begin{eqnarray} \label{eq2}
\frac 1 {NT}\partial_\eta \log Z(\phi)&=& \frac 1 {NT}\meanv{ \sum_{i,t}^T \frac{\mathbb{I}_{g_i^t,g_i^{(t-1)}} - q_{g_i^t}}{\eta \mathbb{I}_{g_i^t,g_i^{(t-1)}} +(1-\eta) q_{g_i^t}}} \\
&=&  \meanv{ \frac 1 {NT}\sum_{i,t}^T \sum_{a,b} \delta_{g_i^ta}\delta_{g_i^{(t-1)}b} \frac{\delta_{a,b} - q_{a}}{\eta \delta_{a,b} +(1-\eta) q_{a}}}\\
&=& \meanv{ \sum_{a=1}^k \frac{1-q_a}{\eta +(1-\eta) q_a} f_{g_i^t=g_i^{(t-1)}=a} - \frac 1 {1-\eta}f_{g_i^t\neq g_i^{(t-1)}}}=0
\end{eqnarray}
where $f_E$ means the empirical frequency of an event $E$ over space and time.  The quantity inside the bracket in Eq. $(\ref{eq2})$ is exactly what we would obtain by fitting a given observed assignment $\boldsymbol{g}$ with a Markov chain. The difference is that now it is averaged over the posterior.  As a first approximation, assuming the posterior to be  peaked around $\hat{\boldsymbol{y}}$, the assignments inferred from the single snapshot procedure,  we can simply find the solution $\eta=\hat{\eta}$ of the polynomial equation
\be\label{eq:le}
 \sum_{a=1}^k \frac{1-q_a}{\eta +(1-\eta) q_a} f_{\hat{y}_i^t=\hat{y}_i^{(t-1)}=a} - \frac 1 {1-\eta}f_{\hat{y}_i^t\neq \hat{y}_i^{(t-1)}}=0.
\ee
Similarly, deriving the log-likelihood with respect to $\xi$, we get 
\begin{eqnarray}\label{eq3}
\frac { 2\partial_\xi \log Z (\phi)}{N(N-1)T} &=& \meanv{\frac {2}{N(N-1)T}\sum_{(i,j)} \sum_{t=1}^T \frac{\delta_{A_{ij}^tA_{ij}^{(t-1)}}  -   p_{g^t_ig^t_j}^{A^t_{ij}}(1-p_{g^t_ig^t_j})^{1-A^t_{ij}}}{\xi\delta_{A_{ij}^tA_{ij}^{(t-1)}}  +(1-\xi) p_{g^t_ig^t_j}^{A^t_{ij}}(1-p_{g^t_ig^t_j})^{1-A^t_{ij}} } }  \\
&=& \meanv{\frac {2}{N(N-1)T}\sum_{(i,j)} \sum_{t=1}^T    \sum_{a,b} \sum_{\epsilon,\epsilon'}^{0,1}  \delta_{g_i^ta} \delta_{g_j^tb}  \delta_{A_{ij}^t\epsilon} \delta_{A_{ij}^{(t-1)}\epsilon'}   \frac{\delta_{\epsilon,\epsilon'}  -  p_{ab}^{\epsilon}(1-p_{ab})^{1-\epsilon}  }{\xi\delta_{\epsilon,\epsilon'}  + (1-\xi)p_{ab}^{\epsilon}(1-p_{ab})^{1-\epsilon}  }}\nonumber\\
&=& \meanv{ \sum_{a,b} \sum_{\epsilon,\epsilon'}^{0,1}  \frac{\delta_{\epsilon,\epsilon'}  -  p_{ab}^\epsilon(1-p_{ab})^{(1-\epsilon)}}{\xi\delta_{\epsilon,\epsilon'}  +(1-\xi)  p_{ab}^\epsilon(1-p_{ab})^{(1-\epsilon)} }m^{ab}_{\epsilon'\to\epsilon} (\boldsymbol{g})}=0
\end{eqnarray}
having the same structure of $(\ref{eq2})$, averaged over the  posterior $(\ref{eq:l2})$ and where we have introduced  the quantities 
\be
m^{ab}_{\epsilon'\to\epsilon}(\boldsymbol{g})=\frac{2}{TN(N-1)}\sum_{t=1}^T\sum_{(i,j)}  \d(g^t_i=a)\d(g^t_j=b)\d(A^{t-1}_{ij}=\epsilon')\d(A^t_{ij}=\epsilon).
\ee 
Again, as soon as the posterior is concentrated around a set of inferred assignments $\hat{\boldsymbol{y}}$, we can simply find the solution $\xi=\hat{\xi}$ of the equation
\be\label{eq:lxi}
\sum_{a,b=1}^k \sum_{\epsilon,\epsilon'}^{0,1}  \frac{\delta_{\epsilon,\epsilon'}  -  p_{ab}^\epsilon(1-p_{ab})^{(1-\epsilon)}}{\xi\delta_{\epsilon,\epsilon'}  +(1-\xi)  p_{ab}^\epsilon(1-p_{ab})^{(1-\epsilon)} }m^{ab}_{\epsilon'\to\epsilon} (\hat{\boldsymbol{g}})=0
\ee
Note that as soon as we use the inferred assignment instead of the full  posterior distribution,  Eqs. ($\ref{eq:le}$) and ($\ref{eq:lxi}$) are not coupled, thus $\hat \xi$ and $\hat \eta$ can be obtained independently.   It is worth noticing that the presence of a temporal lag doesn't affect the result of learning link and group persistences even if we use  $\boldsymbol{y}^t=\hat{\boldsymbol{g}}^{t-\tau^\star}$ instead of $\hat{\boldsymbol{g}}^t$. This is because asymptotically, at large $t$, the lag is constant, thus preserving the ordering,  and the procedure bias can be considered as just a uniform shift over the inferred communities. At the same time,  Eqs.  ($\ref{eq:le}-\ref{eq:lxi}$) work as soon as a sequence of consecutive assignments is considered. In the next subsection we numerically test this procedure to infer the persistence parameters.

\subsection*{Lagged inference.} Starting from the estimates $\hat{\xi}$ and $\hat{\eta}$ we get an estimate of the asymptotic optimal lag as
\be\label{eq:optlag}
\hat{\tau^\star}=\operatorname{argmax}_{\tau} \left( \hat{\xi}^\tau \frac{1-\hat{\xi}}{1-\hat{\xi}\hat{\eta}^2}+(1-\hat{\xi})\hat{\eta}^2\frac{\hat{\eta}^{2\tau}-\hat{\xi}^\tau}{\hat{\eta}^2-\hat{\xi}}\right),
\ee
from which we can shift back the inferred assignments $\hat{\boldsymbol{g}}^{t-\hat\tau^\star}=\boldsymbol{\hat y}^t$ and correct the effective learned assortativity $\hat{a^\star}$ to
\be
\hat{a}=\hat{a^\star}  \left( \hat{\xi}^{\hat{\tau^\star}} \frac{1-\hat{\xi}}{1-\hat{\xi}\hat{\eta}^2}+(1-\hat{\xi})\hat{\eta}^2\frac{\hat{\eta}^{2\hat{\tau^\star}}-\hat{\xi}^{\hat{\tau^\star}}}{\hat{\eta}^2-\hat{\xi}}\right)^{-1}
\ee

\begin{figure}
\includegraphics[scale=0.5]{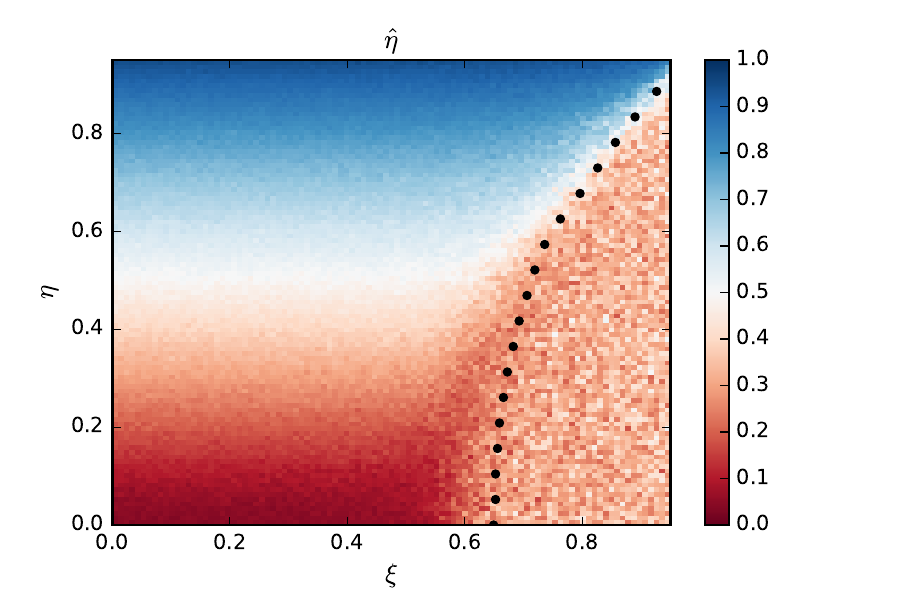}
\includegraphics[scale=0.5]{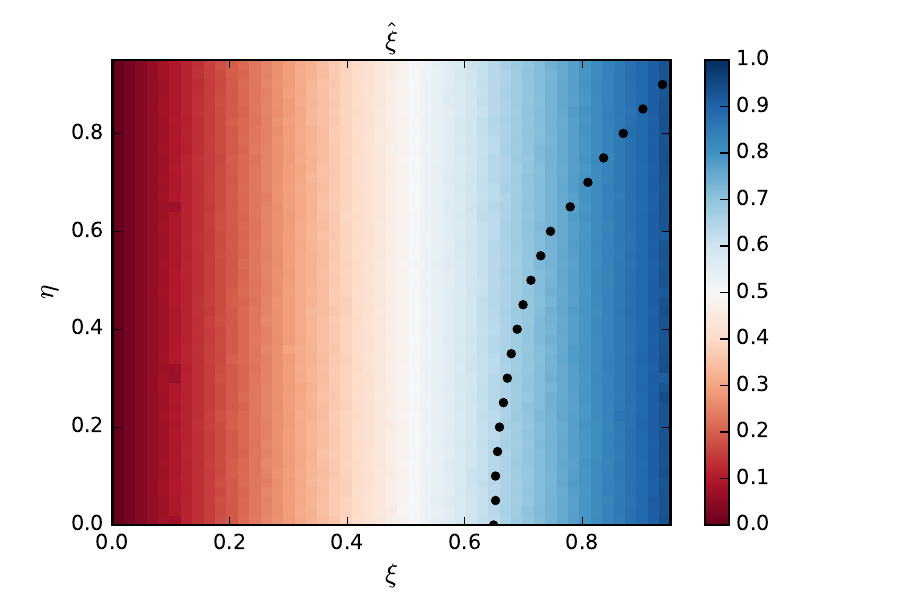}
\caption{Learning $\eta$ and $\xi$ from  synthetic dynamic networks generated according to the DSBM with group and link persistence. We use $T = 50$ snapshots of
networks with $N = 300$ nodes,  $k = 2$ equally sized evolving communities and
planted parameters $\eta$, $\xi$, $a=1.0$, $\bar c = 10$. The panels show the learned $\hat{\eta}$ and $\hat{\xi}$ as function of the planted $\eta$, $\xi$. They coincide at least until the detectability transition line (black dots), where the overlap $q(\hat{\boldsymbol{g}},\boldsymbol{g})$ between inferred and planted assignments vanishes (see top left panel of Fig. \ref{fig:perf}.  }\label{fig:learn}
\end{figure}

We perform extensive numerical simulations to test the effectiveness of the LSD algorithm. Before showing the results of the full LSD, we first test step (2) of the algorithm, which estimates the persistence parameters from the (biased) estimation of the assignments. Fig. \ref{fig:learn}  shows the result of learning $\eta$ and $\xi$ from Eqs. ($\ref{eq:le}$) and ($\ref{eq:lxi}$) using the assignment $\hat{\boldsymbol{y}}$ from  the single snapshot procedure.   The learned parameters $\hat{\eta}$ and $\hat{\xi}$ are in agreement with the planted ones, at least as soon as the overlap between the planted and inferred communities is far from zero.

\begin{figure}
\includegraphics[scale=0.5]{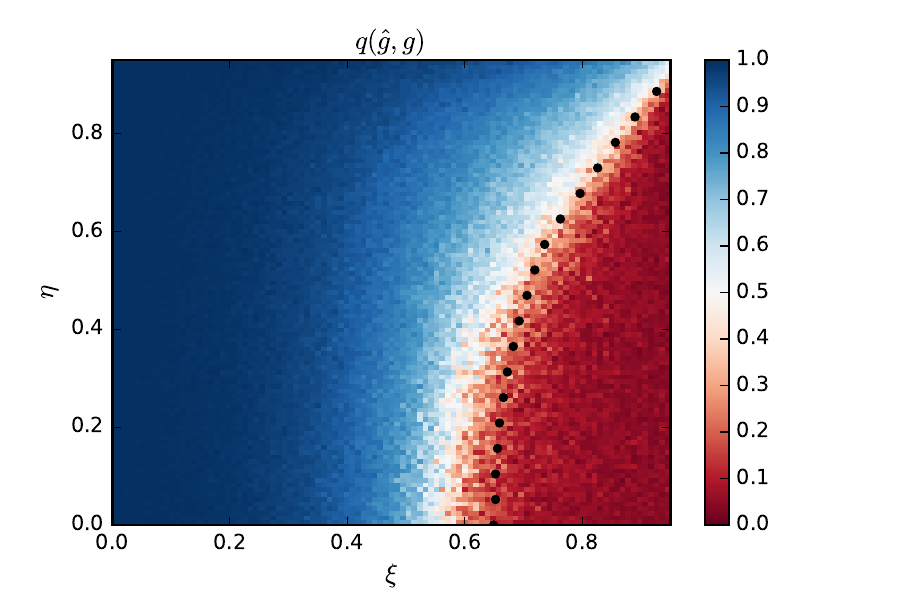}
\includegraphics[scale=0.5]{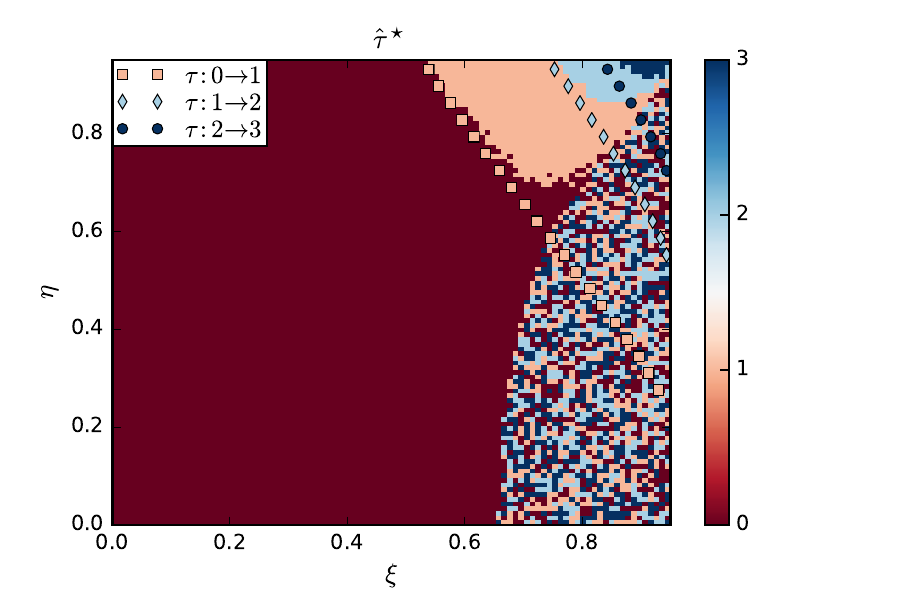}
\includegraphics[scale=0.5]{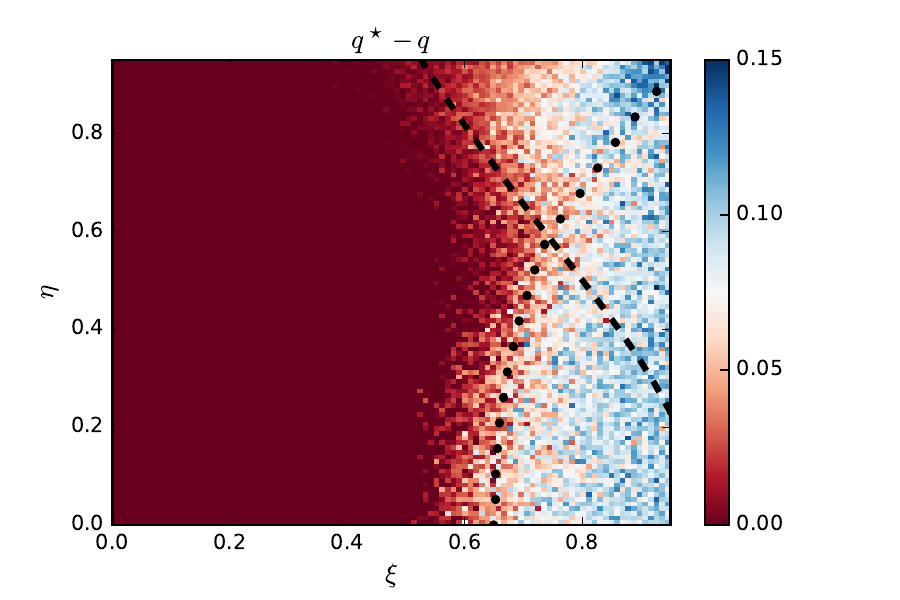}
\includegraphics[scale=0.5]{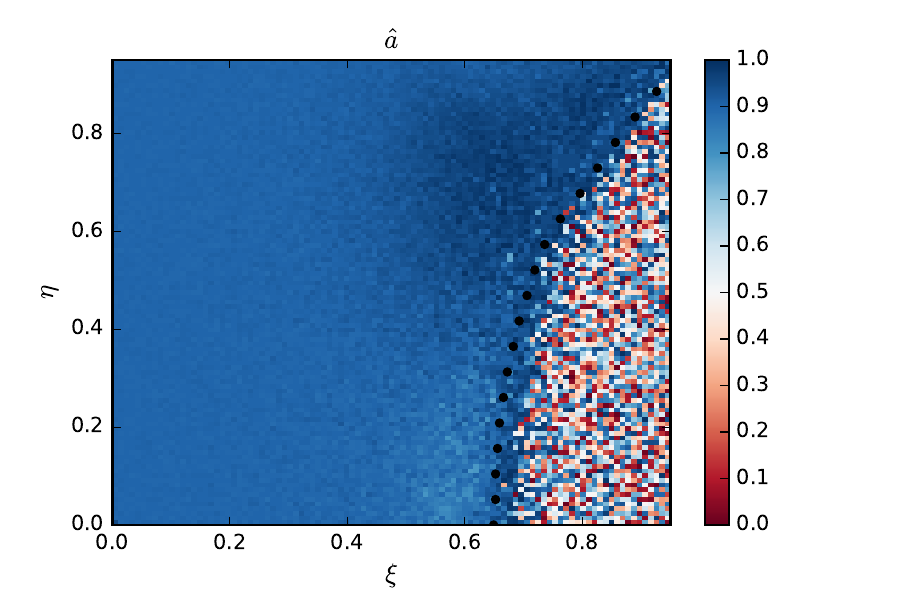}
\caption{Performances of the LSD algorithm on synthetic networks. We use $T = 50$ snapshots of
networks with $N = 300$ nodes,  $k = 2$ equally sized evolving communities and
planted parameters $\eta$, $\xi$, $a=0.9$, $\bar c = 10$. {\it Top left panel:} overlap between the inferred and planted assignments. {\it Top right panel:} optimal inferred lag $\hat \tau^*$ of Eq. \ref{eq:optlag}  and comparison with the analytical transition lines (diamonds) signing lag jumps. {\it Bottom left panel:} difference between $q^\star=q(\hat{\boldsymbol{g}}^t,\boldsymbol{g}^t)$ and $q=q(\boldsymbol{y}^t,\boldsymbol{g}^t)$, i.e. after and before the time lag correction. There is a net positive gain to the right of the dashed line, indicating the first jump from zero to nonzero optimal lag at $\eta^2\xi/(1-\xi)=1$.  {\it Bottom right panel:} learned assortativity $\hat{a}$ as function of $\eta$ and $\xi$.  In all panels black dotted line is the detectability transition line $a^\star_\infty(\xi,\eta)=\bar{c}^{-1/2}$. }\label{fig:perf}
\end{figure}

We then test the performances of the LSD procedure against  synthetic dynamic networks generated according to the DSBM with group and link persistence.  We use $T=50$ snapshots of networks with $N=300$ nodes, mean degree $\bar c=10$,  $k=2$ equally sized evolving communities  and a wide range of planted parameters $\eta$, $\xi$, $a$.  In Fig. \ref{fig:perf} (top left) we show the the overlap  $q(\hat{\boldsymbol{g}},\boldsymbol{g})$ between planted and inferred assignments as a function of $\eta$, $\xi$. For a large region of the parameter space the overlap is very high, showing that the LSD algorithm is able to recover the planted assignment. The black dots indicate the detectability transition line of equation $a^*_\infty(\xi,\eta)=\bar c^{-1/2}$. As expected in the region to the right of this line the overlap is very small. The top right panel shows the estimated value of $\hat \tau^*$ as a function of the persistence parameters. and the top right corner is the region where lagged inference is necessary. In fact the bottom left panel shows $q(\hat{\boldsymbol{g}}^t,\boldsymbol{g}^t)- q(\boldsymbol{y}^t,\boldsymbol{g}^t)$ to highlight the role of time shift in assignment inference. As expected, the region where time shift is critical is the one where $\hat \tau^*$ is  different from zero. The transition line between these two region is described by $\eta^2\xi/(1-\xi)=1$ (dashed line). Finally the bottom right panel shows the inferred $\hat a$, which in the detectability region is always very close to the planted value $a=0.9$.

\section{Comparison with a full dynamic inference}\label{sec:4}

In this Section we compare the LSD algorithm, that is a single-snapshot based algorithm, with a fully dynamic algorithm, i.e. that uses at once the information of the whole time series of network snapshots and the dynamic rules of the generating process. It is a suitable modification of  the dynamic algorithm by Ghasemian et al.  \cite{ghasemian2016detectability}.  This method is based on the observation, by an anonymous referee we want to thank, that if one was able to detect and remove links that have been copied from the past and not generated according to the communities at the present, then it would be as the resulting  temporal network was generated according to the  Ghasemian et al. model, for which their BP algorithm is thought to be optimal. Thus the second method is based on the manual detection and subsequent remotion of links in the snapshot at time $t$ that appears both in the snapshots at time $t$ and $t-1$  (that in the sparse regime are with high probability links that have been copied from the past) followed by the use of the BP algorithm by Ghasemian et al. \cite{ghasemian2016detectability}.  

The comparison in this Section concerns the detectability regions of the two procedures that can be found as suitable modification of the detectability region of a  model without link persistence \cite{ghasemian2016detectability}, being
\be\label{eq:assGha}
a\geq \frac 1 {\sqrt{c}} \sqrt{\frac{1-\eta^2}{1+\eta^2}},
\ee
where $a=(1-\epsilon)/(1+\epsilon)$ is the assortativity parameter, $\epsilon=c_{out}/c_{in}$, and $c$ is the mean degree. In absence of group persistence, we retrieve the static threshold $a\geq 1/\sqrt{c}$.

\subsection*{Case $\eta=0$}
Let's now start by considering the detectability performance of the two methods in the case of a dynamic network with $\xi>0$ and $\eta=0$.   According to the LSD algorithm the role of the link persistence is to decrease the effective assortativity of the network, from $a$ to $a(1-\xi)$. Thus 
\be
a(1-\xi)\geq \frac 1 {\sqrt{c}}  \implies  a\geq \frac 1 {\sqrt{c}(1-\xi)} .
\ee
On the other side, according to the modified Ghasemian method, the network resulting from the cleaning of persistent links  is statistically equivalent to a network generated through the same SBM, but with an effective degree that is smaller, from $c$ to $c(1-\xi)$. Thus the detectability region of this method is expected to be 
\be
a\geq \frac 1 {\sqrt{c(1-\xi)}}=  \frac {\sqrt{1-\xi}} {\sqrt{c}(1-\xi)} 
\ee
that is larger than the one of the LSD algorithm.  The conclusion is that, in both cases, link persistence affects the detectability, reducing the effective assortativity in the first case and the effective degree  in the second case. Both outcomes make the inference harder. Nevertheless detecting and removing the source of the noise, at the cost of reducing the effective degree of the network, is  better than leaving random uninformative links reducing the effective assortativity.

\subsection*{General case $\eta>0$}
In this case the presence of the group persistence mitigates the negative effect of the link persistence. In the LSD algorithm again we need just to consider the effective assortativity  that in this case will be $a(\xi,\eta)$: for example out of the lagged region $a(\xi,\eta)= a(1-\xi)/(1-\xi\eta^2)$. Thus the detectable region will be
\be\label{eq:eq4}
a\frac{1-\xi}{1-\xi\eta^2}\geq \frac 1 {\sqrt{c}} \implies a\geq \frac{1-\xi\eta^2}{(1-\xi )\sqrt{c}}
\ee
that is increasing with $\eta$.  In the lagged region the bias correction allows the detectable region to be a bit larger, just the expression of $a(\xi,\eta)$ is a bit more involved.

\begin{figure}
\includegraphics[scale=0.6]{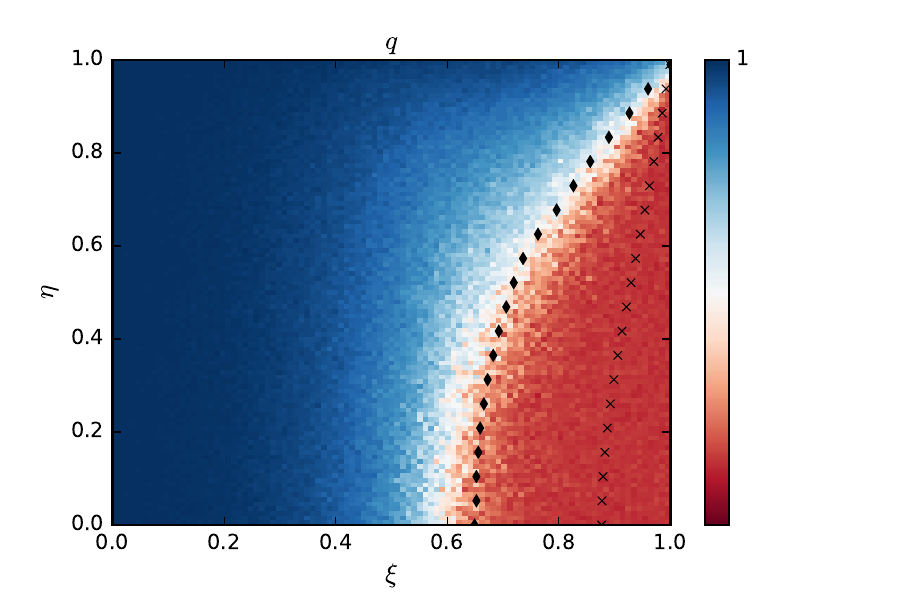}
\caption{Overlap after the LSD inference in terms of the two persistencies ( $T = 50$, $N = 300$,  $a=1.0$, $\bar c = 10$ and  $k = 2$ equally sized evolving communities). Squares and crosses  represent the detectability lines respectively of the LSD algorithm and the full dynamic algorithm based on   \cite{ghasemian2016detectability}. }\label{fig:ph_d}
\end{figure}

In the case of the second method, again we have a decrease in the effective degree but also a net gain in the detectability due to the group persistence and the optimality of the Ghasemian algorithm, i.e. using Eq. (\ref{eq:assGha})
\be\label{eq:eq5}
a\geq \frac 1 {\sqrt{c(1-\xi)}} \sqrt{\frac{1-\eta^2}{1+\eta^2}}.
\ee
Since $1-\xi\eta^2 \geq \sqrt{(1-\eta^2)(1-\xi)/(1+\eta^2)}$, the detectable region of the LSD is always smaller.
To visualize the difference it is possible to plot the critical line in the phase diagram $(\xi,\eta)$, Fig. \ref{fig:ph_d}, keeping fixed $c$ and $a$.  Inverting the previous relations we get respectively for the LSD and the modified Ghasemian method
\be
\xi\leq\frac{\sqrt{c}a -1}{\sqrt{c}a-\eta^2}\ \ \ \ \ \hbox{and}\ \ \  \ \ \xi\leq1-\frac{1}{a^2c}\frac{1-\eta^2}{1+\eta^2}.
\ee
In particular, as soon as $a\geq 1/\sqrt{c}$, both the critical link persistencies tend to 1 when $\eta\to 1$, being maximally different at $\eta=0$. On the contrary if  $a\leq 1/\sqrt{c}$, the LSD algorithm is always in the undetectable region while the Ghasemian one is always in the detectable region for $\eta$ close enough to $1$.   The same comparison can be expressed in terms of the minimal average degree necessary for the detectability: again inverting Eqs. (\ref{eq:eq4}) and (\ref{eq:eq5}) we get, again for LSD and Ghasemian respectively,
\be
c\geq \frac{1}{(1-\xi)a^2}\frac{(1-\xi\eta^2)^2}{(1-\xi)}\ \ \ \ \ \hbox{and}\ \ \ \ \ c\geq \frac{1}{(1-\xi)a^2}\frac{1-\eta^2}{1+\eta^2}.
\ee
We can conclude that both procedures are able to leverage on the group persistence to increase the detectability performances even in the  presence of link persistence (for any $\xi$ the communities are detectable for $\eta$ large enough): the modified Ghasemian algorithm makes use of the whole temporal network, the LSD procedure uses only the information of a single snapshot enriched by the information on past communities codified in the copied links. As expected the LSD procedure cannot be optimal but can be considered as an interesting tradeoff  between a full dynamic inference that uses the whole time series history, accurate but expensive, and a naive single snapshot inference, blind to the dynamics but simpler. Moreover it can be also considered faster and more flexible in the perspective of online inference, in which, given all the information up to $t$, the analysis of a single new snapshot at time $t+1$ is required without examining the entire past. It is faster because in this context the complexity of a full dynamic inference would be $O(TN)$ for each new snapshot, instead of  $O(N)$ for a single snapshot inference and $O(\tau N)$ for LSD with time lag correction.  It is more flexible  because typically real temporal networks are not stationary and a localized-in-time procedure can be desirable for a better tracking of the dynamics.

\section{Conclusions}

We studied the inference problem in a temporal network model where both communities and links are time varying. We focused on static algorithms for temporal networks, where inference is performed on each snapshot network and found that link persistence is the driver of a new kind of detectability transition bringing to  time lagged inference, i.e. a bias towards the detection of  past communities. Analyzing static detection of dynamic communities we were able to introduce a  time-lag corrected procedure, the lagged snapshot dynamic (LSD) algorithm, able to correct the bias thus outperforming naive single snapshot inference. This  algorithm leaves room for improvement from new algorithms that, using the information given by the full temporal network, might reach optimality still maintaining the flexibility of a single-snapshot based approach.  Suitable generalization of a single snapshot based procedure is necessary for online learning techniques where only few  snapshots in the past ($\tau$ in the LSD case), instead of the whole time series, are necessary to get the assignments of a new network snapshot. This is particularly relevant especially if we aim at studying real networks that are far from being stationary. 

\section{Aknowledgment}

Authors acknowledge support from the grant SNS16LILLB - \textit{Financial networks: statistical models, inference, and shock propagation}; PB acknowledges support from FET Project DOLFINS nr. 640772  and FET IP Project MULTIPLEX nr. 317532, from the London Institute for Mathematical Sciences (LIMS) and the University of Zurich (UZH) for providing support during this research collaboration; DT acknowledges support from the grant SNS18ATANTARI - \textit{Dynamic networks: measure, model and mitigate financial risks} and was supported by National Group of Mathematical Physics (GNFM-INDAM).

\bibliographystyle{apsrev4-1}

\begin{thebibliography}{}
\bibitem{holme2012temporal}P. Holme, and J. Saramaki, Physics Reports 519, no. 3: 97-125 (2012).

\bibitem{emid}  P. Mazzarisi, P. Barucca, F. Lillo, D. Tantari, arXiv preprint arXiv:1801.00185 (2018)

\bibitem{newman2016community} M. E. J. Newman, Phys. Rev. E 94, no. 5: 052315 (2016).

\bibitem{hendrickson1995improved}B. Hendrickson, and R. Leland. SIAM Journal on Scientific Computing 16.2: 452-469 (1995).

\bibitem{krzakala2013spectral} F. Krzakala, C. Moore, E. Mossel, J. Neeman,
A. Sly, L. Zdeborov\'{a}, P. Zhang, Proc. Natl. Acad. Sci. USA 110 20935-20940 (2013).

\bibitem{decelle2011asymptotic} A. Decelle, F. Krzakala, C. Moore, L. Zdeborov\'{a}, Phys. Rev. E 84 (6), 066106 (2011).

\bibitem{blondel2008fast}V. D. Blondel \textit{et al.}, Journal of Statistical Mechanics: Theory and Experiment 2008.10: P10008 (2008).

\bibitem{boccaletti} S. Boccaletti {\it et al.}, Physics Reports, 544, 1-122 (2014).


\bibitem{mossel2013proof1}E. Mossel, J. Neeman, and A. Sly. Belief propagation, robust reconstruction and optimal recovery of block models. 
Conference on Learning Theory, 356-370 (2014).

\bibitem{mossel2013proof2}E. Mossel, J. Neeman, and A. Sly, A. Probab. Theory Relat. Fields  162: 431 (2015). 

\bibitem{mucha2010community} P. J. Mucha, T. Richardson, K. Macon, M. A. Porter, and J. Onnela. Science 328, no. 5980: 876-878 (2010).

\bibitem{yang2011detecting}T. Yang, Y. Chi, S. Zhu, Y. Gong, and R. Jin, Machine learning 82, no. 2: 157-189 (2011).

 \bibitem{bassett2013robust}D. S. Bassett, M. A. Porter, N. F. Wymbs, S. T. Grafton, J. M. Carlson, and P. J. Mucha, Chaos: An Interdisciplinary Journal of Nonlinear Science 23, no. 1: 013142 (2013).
 
 \bibitem{bazzi2016community}M. Bazzi, M. A. Porter, S. Williams, M. McDonald, D. J. Fenn, and S. D. Howison, Multiscale Modeling \& Simulation 14, no. 1: 1-41 (2016).

\bibitem{ghasemian2016detectability} A. Ghasemian, P. Zhang, A. Clauset, C. Moore, and L. Peel. Detectability thresholds and optimal algorithms for community structure in dynamic networks. Physical Review X 6.3 (2016): 031005.

\bibitem{zhang2016random} X. Zhang, C. Moore, C., M.E.J. Newman,  Eur. Phys. J. B  90: 200 (2017).

\bibitem{xu2014dynamic}K. S. Xu, and A. O. Hero, IEEE Journal of Selected Topics in Signal Processing 8, no. 4: 552-562 (2014).

\bibitem{xu2015stochastic}K. S. Xu, Stochastic block transition models for dynamic networks. Artificial Intelligence and Statistics (2015).

\bibitem{amaral2000classes}L. A. N. Amaral, A. Scala, M. Barthelemy and H.E. Stanley, Proc. Natl. Acad. Sci. USA 97.21: 11149-11152 (2000).

\bibitem{mossel} E. Mossel, J. Neeman, and A. Sly. A proof of the block model threshold conjecture. Combinatorica (2013): 1-44.

\bibitem{fridman2001elements} J. Friedman, H. Trevor, and R. Tibshirani. The elements of statistical learning. Vol. 1. No. 10. New York, NY, USA:: Springer series in statistics, (2001).

\bibitem{peixoto2013parsimonious} P.T. Peixoto, Physical Review Letters 110, 14: 148701 (2013).

\bibitem{decelle2011inference} A. Decelle, F. Krzakala, C. Moore, L. Zdeborov\'{a}, Phys. Rev. Lett. 107 (6), 065701 (2011).

\bibitem{iba} Y. Iba, Journal of Physics A: Mathematical and General 32, 3875 (1999)

\bibitem{pp1}  M. E. Dyer and A. M. Frieze, J. Algorithm 10, 451 (1989).

\bibitem{pp2}  A. Condon and R. M. Karp, Random Struct. Algor. 18, 116 (2001).

\end{thebibliography}

\end{document}